\documentclass[aps,article,twocolumn,nofootinbib,groupedaddress, showpacs]{revtex4}
\usepackage{graphicx}
\usepackage{amsmath,amssymb}
\usepackage{hyperref}
\usepackage[usenames]{color}

\hypersetup{
    colorlinks=true,
    linkcolor=red,
    citecolor=blue,
}

\def\Journal#1#2#3#4{{#1} {\bf #2}, #3 (#4)}
\def\be{\begin{equation}}
\def\ee{\end{equation}}
\def\ba{\begin{eqnarray}}
\def\ea{\end{eqnarray}}

\newcommand\PRL{Phys.~Rev.~Lett.}
\newcommand\PRD{Phys.~Rev.~D}

\newcommand\etal{{\em et al.}}
\begin{document}

\title{Cosmological tests of general relativity with future tomographic surveys}

\author{Gong-Bo Zhao$^{1}$, Levon Pogosian$^{1}$, Alessandra Silvestri$^{2}$, and Joel Zylberberg$^{3}$}

\affiliation{$^1$Department of Physics, Simon Fraser University, Burnaby, BC, V5A 1S6, Canada\\
 $^2$ Kavli Institute for Astrophysics and Space Research, MIT,
Cambridge, MA 02139, USA \\
$^3$ Department of Physics, University of California, Berkeley, CA
94720, USA
}

\begin{abstract}
Future weak lensing surveys will map the evolution of matter
perturbations and gravitational potentials, yielding a new test of
general relativity on cosmic scales. They will probe the relations
between matter overdensities, local curvature, and the Newtonian
potential. These relations can be modified in alternative gravity
theories or by the effects of massive neutrinos or exotic dark
energy fluids. We introduce two functions of time and scale which
account for any such modifications in the linear regime. We use a
principal component analysis to find the eigenmodes of these
functions that data will constrain. The number of constrained modes
gives a model-independent forecast of how many parameters describing
deviations from general relativity could be constrained, along with
$w(z)$. The modes' scale and time dependence tell us which
theoretical models will be better tested.
\end{abstract}


\maketitle

The observed acceleration of cosmic expansion poses a puzzle for
modern cosmology. It may be evidence for dark energy (DE), a
component with a negative equation of state, $w$, that makes it
gravitationally repulsive. It also warrants studying extensions of
general relativity (GR) with extra degrees of freedom that can mimic
the effects of DE. Modifications to GR are well constrained in dense
regions like our solar system~\cite{Will}. On larger scales,
however, GR is less well-tested. Several modifications to GR,
capable of producing cosmic acceleration have been
proposed~\cite{latest-review}. With the right parameter values, they
can match the expansion history of a universe made of cold dark
matter (CDM) and a cosmological constant $\Lambda$ -- the
observationally favored $\Lambda$CDM model~\cite{wmap5}. However,
their predictions for the growth of structure can differ since the
equations for the evolution of perturbations are modified. Future
tomographic weak lensing surveys, like the Dark Energy
Survey~(DES)\cite{DES} and Large Synoptic Survey Telescope
(LSST)~\cite{LSST}, will measure lensing shear and galaxy counts in
many redshift slices (hence the term tomography), thus mapping the
evolution of perturbations, and offering a new test of GR on
cosmological scales~\cite{latest-review}. In this work, we use a
two-dimensional Principal Component Analysis (PCA) to forecast the
constraints on modified growth (MG) -- and thus our understanding of
gravity -- coming from these surveys. Unlike previous MG forecasts,
ours is model-independent, and lets us determine how many parameters
describing MG could be constrained, along with the regions in
parameter space where we expect the most sensitivity to MG.

We consider linear scalar perturbations to the flat
Friedmann-Robertson-Walker metric in Newtonian gauge
\begin{equation}\label{FRW}
ds^2=-a^2(\eta)[(1+2\Psi(\vec{x},\eta))d\eta^2-(1-2\Phi(\vec{x},\eta))d\vec{x}^2],
\nonumber
\end{equation}
where $\eta$ is the conformal time and $a(\eta)$ the scale factor.
In Fourier space, one can write~\cite{Hu:2007pj,BZ08}
\begin{eqnarray}\label{gamma}
\Phi/\Psi=\gamma(k,a), \ \ \label{parametrization-Poisson}
k^2\Psi=-\mu(k,a) 4\pi G a^2\rho\Delta \ ,
\end{eqnarray}
where $\Delta$ is the comoving matter density perturbation. The
function $\gamma$ describes anisotropic stresses, while $\mu$
describes a time- and scale-dependent rescaling of Newton's constant
$G$, as well as the effects of DE clustering (a feature of many
exotic DE models) or massive neutrinos. In $\Lambda$CDM, the
anisotropic stress due to radiation is negligible during matter
domination, thus $\gamma=1=\mu$. In this letter, we determine how
well the unknown functions $\gamma(k,a)$ and $\mu(k,a)$ can both be
constrained by future data.  We also address how well we can detect
any departure from $\gamma=1=\mu$, without distinguishing between
them.

We consider the two-point correlations (both auto- and cross-)
between galaxy counts (GC), weak lensing shear (WL), and cosmic
microwave background (CMB) temperature anisotropy, plus the CMB
E-mode polarization and its correlation with the CMB temperature.
Detailed descriptions of our assumptions for each measurement are
found in~\cite{ZPSZ08}. GC probe the distribution and growth of
matter overdensities, thus giving an estimate of the Newtonian
potential $\Psi$, up to a bias factor. WL is sourced by the sum of
the potentials $(\Psi+\Phi)$. GC are also affected by
``magnification bias''~\cite{bias}, where WL convergence magnifies
some faint (thus otherwise undetected) galaxies, adding mild
dependence on $(\Psi+\Phi)$ to the GC. CMB data probe the Integrated
Sachs-Wolfe effect (ISW) which depends on $d(\Phi+\Psi)/d\eta$.
Thus, measuring GC and WL over multiple redshift bins, along with
CMB data, yields information about the relation between $\Psi$ and
$\Phi$ and their response to matter density fluctuations.
Furthermore, supernovae (SN) redshift-luminosity measurements and
CMB constrain the expansion history. For our forecasts, we assume
the following probes: Planck~\cite{Planck} for CMB, DES and later
LSST for GC and WL, and a Supernova/Acceleration Probe -like the
Joint Dark Energy Mission (JDEM)~\cite{JDEM} for SN. To compare to
current data, we use large scale structure (LSS) data compiled
in~\cite{Giannantonio:2008zi}, and the latest SN and CMB data from
the `Constitution' SN sample~\cite{CfA} and WMAP5~\cite{wmap5}. For
current data, we omit WL, but include the GC-CMB cross-correlations.

In full generality, we treat $\gamma(k,a)$ and $\mu(k,a)$ as unknown functions and perform a PCA~\cite{tegmark-pca,PCA1} to determine how many of their d.o.f. can be constrained. We use redshift  $z\equiv a^{-1}-1$ as a time variable, and pixelize the late-time and large-scale universe ($z\in[0,30], k\in[10^{-5}, 0.2]~{\rm
h}\,{\rm Mpc}^{-1}$) into $M+1$ $z$-bins and $N$ $k$-bins, with each of the $(M+1)\times N$ pixels having independent values of
$\mu_{ij}$ and $\gamma_{ij}$. We consider $w(z)$ as another unknown function, allowing each of the $M+1$ $z$-bins to have an independent value of $w_i$. Since the surveys we consider will not probe $z>3$ in detail, we use $M$ bins linear in $z$ for $z\in[0,3]$ and a single bin for $z\in[3,30]$. We choose $M=N=20$ and have checked that this pixelization is fine enough to ensure the convergence of the results. We use logarithmic $k$-bins on superhorizon scales and linear $k$-bins on subhorizon scales, to optimize computational efficiency. As in~\cite{ZPSZ08}, we only consider information from scales well-described by linear perturbation theory, which is only a fraction of the $(k,z)$-volume probed by future surveys. Since the evolution equations~\cite{ZPSZ08} contain time-derivatives of $\mu(k,z)$, $\gamma(k,z)$ and $w(z)$, we follow~\cite{PCA2} and use hyperbolic tangent functions to represent steps in these functions in the $z$-direction, while steps in the $k$-direction are left as step functions.

Our parameters are the {$\gamma_{ij}$},  {$\mu_{ij}$}, and {$w_i$},
along with the usual cosmology parameters: energy density of baryons
$\Omega_b{h}^{2}$ and CDM $\Omega_c{h}^{2}$, Hubble constant $h$,
optical depth $\tau$, spectral index $n_s$ and amplitude $A_s$. We
include one bias parameter per GC $z$-bin, and the intrinsic SN
magnitude. Thus we have $(M+1)(2N+1)+17=878$ parameters in total.
For a given set of parameter values, we use
MGCAMB~\cite{ZPSZ08,mgcamb}, (a modification of CAMB \cite{camb}
developed by us to study modified growth), to compute angular
spectra for our observables. We generate numerical derivatives of
observables with respect to parameters, and use the specifications
for the experiments to compute the Fisher information matrix, which
defines the sensitivity of the experiments to these parameters
(see~\cite{ZPSZ08} for computational details). Our fiducial values
are in all cases $\Lambda$CDM: $\gamma_{ij} = \mu_{ij} = -w_i =
1~\forall~i,j$, and the fiducial values of the other parameters are
those of WMAP5~\cite{wmap5}.

Let us first study the expected errors on $\mu(k,z)$. The error on
any $\mu_{ij}$ is large, and the pixels have highly correlated
errors. PCA finds the linear combinations of pixels with
uncorrelated errors. We take only the $\mu_{ij}$ block of the
covariance matrix, thus marginalizing over all other parameters,
including the {$w_i$} and {$\gamma_{ij}$}. We invert this block to
obtain the Fisher matrix for our $\mu$ values, $F_{(\mu)}$, and
diagonalize $F_{(\mu)}$ by writing $F_{(\mu)}=W^{T}\Lambda{W}$. The
rows of matrix $W$ are the eigenvectors, or the Principal Components
(PC's)~\cite{tegmark-pca}, while the diagonal elements of $\Lambda$
are the eigenvalues $\lambda_m$. Each eigenvector, $e_\mu(k,z)$, is
a linear combination of the original pixels $\mu_{ij}$, forming a
surface in $(k,z)$ space. The eigenvectors are orthogonal. We
normalize them to unity, rescaling the eigenvalues accordingly.
Then, $\{e_\mu(k,z)\}$ forms an orthonormal basis in which we can
expand $\mu$ as $\mu(k,z)-1=\sum_{m}\alpha_{m}e_{m}(k,z)$, where
$\alpha_m$ are the new uncorrelated parameters with variances given
by the $\lambda_m$: $\lambda_{m}=[\sigma^{2}(\alpha_{m})]^{-1}$. We
expect, from existing data, that variations in $\mu$ larger than
$\mathcal{O}(1)$ are unlikely. We enforce this by applying a prior
$\lambda_m>1$ to the matrix $F_{(\mu)}$. This procedure, analogous
to the treatment of $w(z)$ in~\cite{JDEMFOM}, does not affect the
well-measured modes, but gives a reference point with respect to
which we define poorly constrained modes. The worst-measured modes
have variances approaching the prior, while those with smaller
variances are the well-measured ones. Since we compute the full
covariance matrix, then marginalize over all but the parameter(s) of
interest, our procedure yields the results that we would get for
$\mu$ if we simultaneously measured $w$, $\gamma$, and $\mu$. This
analysis can be repeated for $\gamma$ or $w$. Given the PCA results,
one can convert uncertainties in the expansion parameters $\alpha_m$
into uncertainties in any other parameterization of $\mu$($\gamma$)
without recalculating the Fisher matrices~\cite{PCA1,PCA2}: one
projects the PC Fisher matrix onto a new basis.

\begin{figure}[tbp]
\includegraphics[scale=1.8]{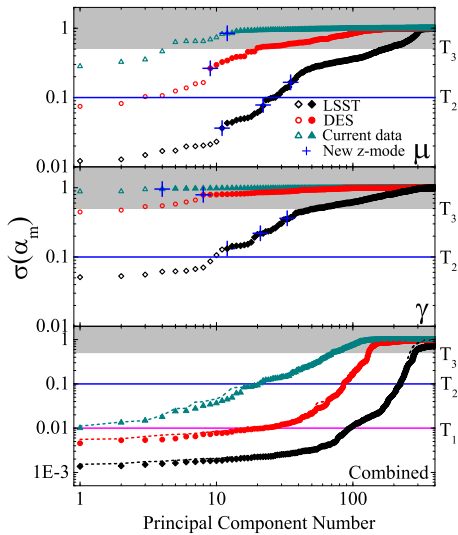}
\caption{Uncertainties in the eigenmodes for
current data, and for future data sets including LSST(DES). The two upper panels show modes of $\mu$ and $\gamma$ with mutual marginalization. The lower panel shows uncertainties in the combined modes. The purple and blue solid lines and the shaded region denote the thresholds $T_1, T_2$
and $T_3$, respectively. The filled symbols denote the redshift-dependent modes. In the lower panel, the dashed lines show the uncertainties for $\mu$ with $\gamma$ fixed.} \label{fig:evalue}
\end{figure}

Measurements probe combinations of $\Phi$ and $\Psi$, so the effects
of $\gamma$, which affects only $\Phi$, are mixed with those of
$\mu$, which affects both potentials. This yields degeneracy between
$\mu$ and $\gamma$. By varying both, then marginalizing over one, we
lose information common to both functions. This is necessary when
separately constraining $\mu$ and $\gamma$. While this degeneracy
impedes their ability to do so, DES and LSST will yield non-trivial
constraints on $\mu$ and $\gamma$ with mutual marginalization
(marginalizing over $\mu$ when measuring $\gamma$, and vice versa).
The uncertainties associated with PC's of $\mu$ and $\gamma$ are
shown in the top two panels of Fig.~\ref{fig:evalue} for LSST, DES,
and current data. To quantify the sensitivity of the surveys to MG,
we introduce three thresholds:  well-constrained ($T_1$,
$\sigma(\alpha_m)\lesssim 0.01$), constrained ($T_2$, $0.01
\lesssim\sigma(\alpha_m)\lesssim 0.1$), and informative ($T_3$,
$0.1\lesssim\sigma(\alpha_m)\lesssim 0.5$). From
Fig.~\ref{fig:evalue}, we see that DES could constrain two $\mu$
parameters and no $\gamma$ parameters. LSST could constrain many
modes, as it will have a superior sky coverage and resolution, wider
z-span, and more precise photometric redshift measurements. Current
data effectively cannot constrain either $\mu$ or $\gamma$. The
constraints on $\mu$ are generally stronger than those on $\gamma$:
$\mu$ affects GC, WL, and CMB, while $\gamma$ primarily affects WL
and CMB (GC is only affected by $\gamma$ via magnification bias).

\begin{figure}[tbp]
\includegraphics[scale=2]{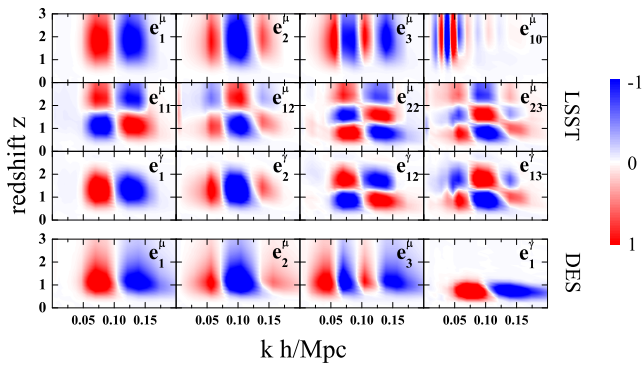}
\caption{Eigensurfaces for $\mu$ and $\gamma$, with mutual marginalization, for LSST(DES) along with Planck and JDEM.} \label{fig:evec1}
\end{figure}

Fig.~\ref{fig:evec1} shows selected eigenmodes of $\mu$ and $\gamma$ for LSST and DES. The $i^{th}$ $\mu$($\gamma$) mode represents the $i^{th}$ best-constrained independent $\mu(k,z)$($\gamma(k,z)$) surface. Models that predict $\mu$ and $\gamma$ similar to our ``best" modes, then, will be better constrained. We observe no degeneracy in the $k$ and $z$ dependences of the modes. This is counter-intuitive, since changing $\mu$ at some point $(k,z)$ should have the same impact on the observables as a change at a larger scale but later time. However, since we allow for simultaneous variation of $\gamma$, the change in $\mu$ is more efficiently off-set by adjusting $\gamma$, eliminating the $k$--$z$ degeneracy. There is a clear pattern to the modes; the best constrained modes have no $z$ nodes, but apparent $k$ nodes, and, approximately, the $m^{th}$ mode has $m$ $k$ nodes. For LSST, at roughly the $10^{th}$ mode, the first $z$ node appears, followed by another period of scale-dependent patterns. The alternating $(k,z)$ patterns repeat until mixed $(k,z)$ modes appear, after which there are no clear patterns. The best modes are mainly functions of $k$ and not $z$. This is partly because the total observable volume in the radial ($z$) direction is limited by the dimming of distant objects and, ultimately, the fact that structures only exist at relatively low $z$. Also, it is related to us considering only linear perturbations in our analysis, since at small $z$ the observable volume is too small to fit the small $k$-modes that are still in the linear regime. Hence, there is more volume available for studying the spatial distribution of structure than the radial distribution. The number of nodes in the $z$ and $k$ directions tell us, respectively, the number of $z$- and $k$-dependent parameters that
surveys could constrain. For example, for LSST, there are three clear $z$-dependent patterns of $\mu$ and $\gamma$ whose eigenvalues fall within $T_2$ and $T_3$ ranges. For DES, the only constrained modes are those of $\mu$, and exhibit only one type of $z$-dependence, with one node.

The LSST modes for $\mu$ and $\gamma$ have similar shapes except that the $\mu$ modes have a deeper $z$ span than the $\gamma$ modes. This is due to an accumulation effect on $\mu$.  On subhorizon scales, the density contrast, (related to $\Psi$ via Eq.~(\ref{parametrization-Poisson})), evolves via $\ddot\Delta+\mathcal{H} \dot\Delta = 4 \pi G \mu\rho{a^2}\Delta$. Perturbing $\mu$ at one pixel, e.g. enhancing it at some $k$ and $z= 3$, enhances $\Delta$ (and thus $\Psi$) for that $k$-mode $\forall~z<3$, since the growth factor at all later times is enhanced. On the other hand, changes in $\gamma$ at high $z$, which primarily affect WL through changes in $\Phi$ (Eq.~(\ref{gamma})), do not affect WL at low $z$. Hence, the $z$-sensitivity range of $\gamma$ is primarily determined by the redshift range of the WL kernel. The $z$-dependence of  $\gamma$ also affects the ISW contribution to the CMB at small $k$ and low $z$, but its contribution to the Fisher matrix is small due to a large cosmic variance.

In most models of modified gravity, $\mu$ and/or $\gamma$ evolve in a time- and scale-dependent way~\cite{latest-review}. For example, in scalar-tensor theories, they undergo a step-like transition at the Compton scale of the scalar field. The peaks in the eigensurfaces indicate the ``sweet spots'' in the $(k,z)$ space where such a transition scale can be detected by the survey, while the frequencies of the modes tell us how well a transition can be resolved. For LSST, the best-measured modes peak in the region $0.04<k<0.16\,{\rm h^{-1}Mpc}$  and $0.5<z<2$, indicating a sensitivity to a Compton scale today of $50\lesssim \lambda_c^0\lesssim 1500\,{\rm Mpc}$, where we have allowed for a range of possible time evolutions of the mass scale\cite{ZPSZ08}. Massive neutrinos also introduce a transition in $\mu$ due to free-streaming. From the expression for the free streaming length in terms of $z$ and the neutrino mass $m_\nu$ (see e.~g. \cite{Pastor}), we find that the transition scale is within the LSST sensitivity window for $ 0.1\lesssim m_\nu \lesssim 0.7\,{\rm eV}$. Smaller masses induce an overall suppression of growth with no scale-dependent signatures. While observable to some extent, this suppression is largely degenerate with $w(z)$, especially if one allows for an arbitrary evolution of $w(z)$, as we have done.

In addition to constraining $\mu$ or $\gamma$ individually, a less
ambitious yet equally interesting question is how sensitive data is
to any departure from standard growth. Namely, one may ask if either
function deviates from unity, without specifying which. For this
purpose, we want to save the information common to both functions,
which we previously lost by mutual marginalization. Hence, we
consider the combined principal components of $\mu$ and $\gamma$. We
follow the same procedure as before, except now we diagonalize the
block of the Fisher matrix containing $\mu$ and $\gamma$ pixels. The
eigenvalues of these combined PC's are shown in the lower panel of
Fig.~\ref{fig:evalue}. Even today's data can provide around $15$
``constrained'' modes. This is unsurprising since the LSS power
spectrum, $P(k)$, is known to better than $10$\% precision over an
order of magnitude in $k$. Changing $\mu$ at any $k$ directly
affects $P(k)$, which means that large variations in $\mu$ are
disallowed in the range where $P(k)$ is well-measured. After
marginalizing over $\gamma$, the direct impact on $P(k)$ is lost,
since one can now offset changes in $\mu$ by adjusting $\gamma$. The
dashed lines in Fig.~\ref{fig:evalue} represent the eigenvalues of
$\mu$ without marginalizing over $\gamma$. They are comparable to
the eigenvalues of combined PC's, supporting the above notion. From
Fig.~\ref{fig:evalue}, we see that DES will provide around $20$,
while LSST will provide up to $100$, ``well-constrained" combined
modes.

So far, our analysis has neglected systematic errors, which are
model-dependent and very hard to predict. While we address
systematics thoroughly in an upcoming publication, we report here
the outcome of a preliminary analysis based on the assumptions
in~\cite{systematicsDH,systematicsZ}. We repeat our PCA with extra
parameters describing likely sources of systematic error: shifts in
the centroids of the $z$-bins, distortions of the $z$-bin
distribution functions, and additive and multiplicative errors on
the WL signal due to point-spread-function contributions, as
in~\cite{systematicsDH}. This adds $58$($80$) parameters to our
analysis for DES(LSST). We assume no ``catastrophic" photo-$z$
mis-estimation, and apply a conservative set of
priors~\cite{systematicsZ} to these parameters and marginalize over
them. We find that the systematics result in a noticeable, but not a
dramatic, dilution of constraints on MG from DES. This is because
photo-z errors would most immediately affect the z-dependence of MG,
to which DES was only weakly sensitive even without the systematics.
As discussed above, constraints from DES will be primarily on the
scale-dependence of $\mu$ and $\gamma$, and that information is
mostly preserved. The impact of the systematics on LSST forecasts is
more significant, because LSST has a higher potential for resolving
z-dependent features. There too we find that inclusion of systematic
errors preserves most of the scale-dependent information but can
reduce our ability to measure eigenmodes of $\mu$ with z-dependent
features, underscoring the need to study and control systematic
errors in lensing surveys. Even after accounting for systematics,
LSST and DES are powerful probes of MG.

The Dark Energy Task Force~\cite{DETF} recently analyzed constraints
on DE from future surveys, without considering MG, and found that
$w(z=0)$ and $(dw/dz)|_{z=0}$ could both be constrained. A
time-varying $w(z)$ alters the growth dynamics in a
scale-independent way, so the scale-dependence of $\mu$ and $\gamma$
cannot be duplicated by a choice of $w$. Furthermore, since we
consider linear scales, the dominant portion of the information on
MG comes from higher $z$ ($z>0.5$), at which DE effects are not as
important. Thus, in addition to measuring $w(z)$, future surveys
will tightly constrain scale-dependent departures from $\Lambda$CDM,
and those occurring at high redshifts. Note that including the
non-linear growth data from lower $z$ requires a model dependent
treatment of MG, in which case $w(z)$ and MG would be related by the
same theory.

To recap, we have forecasted the constraints on modifications to GR from future data sets, in comparison with existing data. We find that combined data from Planck, JDEM and LSST(DES) can tightly constrain around 100(20) parameters of MG (corresponding to the 100(20) eigenvalues in region $T1$ in the combined $\mu$ and $\gamma$ analysis), if the systematics are negligible. Current data can constrain only one parameter to this level. We have further identified the regions in parameter space to which future datasets are most sensitive. In general, our technique can be used in survey design to move the sweet spots to the most interesting parts of parameter space. Our results are obtained using only linear-scale data, being a conservative ``proof of concept'' that upcoming surveys can rigorously test GR over cosmic distances.

{\it Acknowledgments} We thank R. Crittenden, C. Shapiro, E.
Bertschinger, T. Giannantonio, and K. Koyama for useful comments and
discussion. L. P. and G. B. Z. are supported by the NSERC and SFU,
A. S. by the NSF Grant No. AST- 0708501, and J. Z. by the Fulbright
Foundation, NSERC, and UCB.

\end{document}